\documentclass[aps,prx,twocolumn,superscriptaddress,floatfix]{revtex4-2}
\usepackage[dvipsnames,svgnames]{xcolor}
\usepackage{adjustbox}
\usepackage{graphicx}
\usepackage{dcolumn}
\usepackage{bm}
\usepackage{soul}
\usepackage[T1]{fontenc}
\usepackage{amstext}
\usepackage{amsthm}
\usepackage{amsmath}
\usepackage{amsfonts}
\usepackage{amssymb}
\usepackage{appendix}
\usepackage{multirow}
\usepackage{physics}
\usepackage[bookmarks,bookmarksnumbered]{hyperref}
\usepackage{bookmark}
\hypersetup{colorlinks,allcolors=black}
\usepackage{enumitem}
\usepackage{mathtools}
\usepackage[text]{esdiff}
\usepackage{setspace}
\usepackage[caption=false]{subfig}
\usepackage{xspace}
\usepackage{soul}
\usepackage{comment}
\usepackage[utf8]{inputenc}
\usepackage{algpseudocode}

\newcounter{algorithmctr}

\begin{document}
\preprint{APS/123-QED}

\title{Robust Parametric Quantum Gate Against Stochastic Time-Varying Noise}

\author{Yang He}
\author{Zigui Zhang}
\author{Zibo Miao}
\email{miaozibo@hit.edu.cn}
\affiliation{School of Intelligence Science and Engineering, Harbin Institute of Technology, Shenzhen, Shenzhen 518055, China}
\date{\today}

\begin{abstract}

The performance of quantum processors in the noisy intermediate-scale quantum (NISQ) era is severely constrained by environmental noise and other uncertainties.
While the recently proposed quantum control robustness landscape (QCRL) offers a powerful framework for generating robust control pulses for parametric gate families, its application has been practically restricted to quasi-static noise.
To address the spectrally complex, time-varying noise prevalent in reality, we propose filter function-enhanced QCRL (FF-QCRL), which integrates filter function formalism into the QCRL framework.
The resulting FF-QCRL algorithm minimizes a generalized robustness metric that faithfully encodes the impact of stochastic processes, enabling robust pulse-family generation for parametric gates under realistic time-varying noise. Numerical validation in a representative single-qubit setting confirms the effectiveness of the proposed method.

\end{abstract}
\maketitle


\section{Introduction}
The fidelity and robustness of quantum gates are critically limited by environmental noise in the noisy intermediate-scale quantum (NISQ) era, presenting a major obstacle to scalable quantum computing \cite{NISQ_2018}.
To mitigate the impact of noise on computational performance, numerous strategies have been proposed, including quantum error correction codes \cite{QEC_2015} and robust quantum control techniques \cite{QROC_2022}.
Unlike error correction, which operates at the circuit or logical level, robust quantum control focuses on enhancing the resilience of individual quantum gates to noise while achieving target operations.
Established methods such as composite pulses \cite{BB1_SK1_2004,UCPs_2014}, geometric methods \cite{zeng_geo_2019,RQG_geo_2024}, and sampling-based methods \cite{bGRAPE_2019,dong_SLC_2019,smROC_2024} have demonstrated success in specific scenarios.
However, their practical utility is often limited by stringent experimental constraints, including pulse duration, bandwidth limitations, and maximum power thresholds.
Furthermore, these algorithms require re-optimization when operational settings change—for instance, when shortening pulse durations or adapting to new control targets—introducing significant computational overhead.

To address these limitations, the recently proposed quantum control robustness landscape (QCRL) framework \cite{QCRL_2024}, derived from the quantum control landscape (QCL) theory \cite{QOCTL_2004,levelset_QCL_2006,uni_gen_QCL_2008,time_optimal_2015}, offers a complementary approach to existing robust control methods.
Traditional QCL studies focus on mapping control parameters to objective metrics such as gate fidelity, whereas QCRL explicitly characterizes the relationship between control parameters and robustness against noise.
A key application of QCRL lies in its ability to generate optimized solutions from an initial set of control pulses while preserving both fidelity and robustness almost invariant.
This capability is particularly valuable for designing parameterized quantum gates. 
However, a significant limitation of current QCRL implementations is their predominant reliance on quasi-static or deterministic time-varying noise models, which fail to capture the stochastic, spectrally complex noise prevalent in real-world quantum systems.
This gap between the model and the physical reality constrains the practical utility of the framework.

In this work, we extend the QCRL framework to address time-varying stochastic noise by integrating the filter function formalism \cite{FF_2013, CPs_FF_2014, RQG_FF_2017, Qctrl_FF_2021}, a powerful tool for quantifying the impact of stochastic noise processes on quantum gate performance.
We term this extended framework \emph{filter function-enhanced QCRL} (FF-QCRL).
The filter function formalism enables the evaluation of robustness by mapping noise spectra and control parameters to gate infidelity, thereby serving as a robustness metric for QCRL optimization.
We demonstrate the utility of this enhanced framework through a representative single-qubit parameterized-gate setting, where control pulses are systematically generated to maintain robustness against stochastic noise while achieving target gate parameters.
As an auxiliary quantum-assisted implementation route, we additionally introduce a variational quantum circuit (VQC) surrogate for approximating the robustness metric during RIPV-based pulse-family generation, in line with recent surrogate-modeling practice for noisy and data-limited settings \cite{icaart24}.

The remainder of this paper is organized as follows.
Section 2 briefly introduces the concepts of robust quantum control, parametric quantum gates, QCRL, and filter function formalism.
In Section 3, we present the details of the proposed method, including the integration of filter function formalism into the QCRL framework.
Section 4 introduces a variational quantum circuit surrogate for robustness evaluation in the RIPV stage.
In Section 5, we present a representative numerical study of the proposed method for the quantum gate $R_x(\theta)$ under stochastic detuning noise.
Finally, conclusions are drawn in Section 6.

\section{Preliminaries and Problem Formulation}

We begin by establishing the mathematical framework for robust quantum control and then introduce the key tools—QCRL and filter function formalism—that will be integrated in our proposed method.

\subsection{Robust Quantum Control}

The Hamiltonian for the dynamics of the closed quantum system can be written as
\begin{align} \label{hamiltonian_total}
    H_{\rm tot}(t)&=H_{\rm ctrl}(t)+H_{\rm noise}(t)\nonumber\\
    &=H_0+\sum_{j=1}^{N_u} u_j(t) H_{c,j}+\sum_{k=1}^{N_n} \epsilon_k(t) H_{n,k},
\end{align}
where $H_{\rm ctrl}(t)$ is the deterministic part that includes the free Hamiltonian $H_0$ and the sum of the control Hamiltonian $H_{c,j}$ with the corresponding control pulse $u_j(t)$ ($j = 1, \ldots, N_u$, where $N_u$ is the number of control channels), and $H_{\rm noise}(t)$ characterizes the influence of noises that can be viewed as the sum of the noise Hamiltonian $H_{n,k}$ with the corresponding time-dependent noise strength $\epsilon_k(t)$ ($k = 1, \ldots, N_n$, where $N_n$ is the number of noise channels). Using the interaction picture in the control frame $U_{\rm ctrl}(t)$, the propagator of noisy dynamics can be given by
\begin{align}
    U_{\rm tot}(t)&=\widetilde U_{\rm noise}(t)U_{\rm ctrl}(t)\nonumber\\
    &=\mathcal{T}\exp\left(-i\int_0^t\widetilde H_{\rm noise}(\tau)\dd \tau\right)U_{\rm ctrl}(t),
\end{align}
where $\mathcal T$ is the time-ordering operator, and the error Hamiltonian in the interaction picture can be expanded as
\begin{align}
    \widetilde H_{\rm noise}(\tau)&=\sum_{k=1}^{N_n} \epsilon_k(\tau)U_{\rm ctrl}^\dagger(\tau)H_{n,k} U_{\rm ctrl}(\tau)\nonumber\\
    &=\sum_{k=1}^{N_n}\epsilon_k(\tau)\widetilde H_{n,k}(\tau).
\end{align}

Given the target time $T$ and the target operator $U_{\rm target}$, the robust quantum control problem can be formulated as follows: find a control pulse such that $U_{\rm ctrl}(T) \approx U_{\rm target}$ while simultaneously ensuring that the noise-induced evolution $\widetilde U_{\rm noise}(T)$ remains close to the identity operator $\mathbb{I}$.
A typical objective function (infidelity) measuring the distance between two unitary operators $U_1,U_2$ up to an arbitrary global phase factor in some subspace of interest is defined as 
\begin{equation} \label{infidelity_ideal}
    \mathcal{I}(U_1,U_2)=1-\left|\frac1{{\rm Tr}(P)}{\rm Tr}\left(PU_1^\dagger U_2\right)\right|^2,
\end{equation}
where $P$ is the projection operator related to the subspace of interest.
For example, a trivial choice of $P$ is the identity operator $\mathbb{I}$.
Therefore, a robustness performance index can be obtained by taking the expectation of infidelity between $\widetilde U_{\rm noise}(T)$ and $\mathbb {I}$:
\begin{equation}
    \mathcal{I}_{\rm robust}=1- \mathbb{E}\left[\left|\frac1{{\rm Tr}(P)}{\rm Tr}\left(P\widetilde U_{\rm noise}(T)\right)\right|^2\right],\label{eq:robust_eq}
\end{equation}
where $\mathbb{E} \left[ \,\cdot\, \right]$ denotes the expectation over all possible noises. 

\subsection{Parametric Quantum Gate}

The parametric quantum gate is a class of quantum operations in quantum computing that are characterized by tunable parameters. For $n$ qubits, let $\boldsymbol{\theta}=(\theta_1,\theta_2,\cdots,\theta_m)\in \mathbb{R}^m$ denote a set of parameters, a parametric quantum gate can be defined as a mapping from the parameter space to the set of unitary matrices, that is, $U(\boldsymbol{\theta}):\mathbb{R}^m\mapsto \mathcal{U}(2^n)$. Representative examples include the rotating operators $R_{\{x,y,z\}}(\theta)$ which rotate a single qubit along the $x,y,z$ axes by an angle $\theta$; controlled rotating operators $|c\rangle |t\rangle\mapsto |c\rangle R_{\{x,y,z\}}^c(\theta)|t\rangle$  that rotate the target qubit when the control qubit is set.

This tunability endows parametrized quantum gates with greater flexibility and adaptability, making them particularly useful in the design of quantum algorithms and the construction of quantum circuits for a wide range of complex quantum computing tasks.

\subsection{Quantum Control Robust Landscape}

The quantum control robustness landscape (QCRL) is a theoretical framework that maps control parameters to the robustness of quantum operations against noise, rather than focusing solely on fidelity.
Unlike the conventional quantum control landscape (QCL), QCRL offers a new approach to evaluate and optimize the performance of quantum gates and control pulses under noisy conditions.
Mathematically, QCRL is a map from control parameters $\boldsymbol{c}$ to some robustness measure $R(\boldsymbol{c})$, that is, $\boldsymbol{c} \mapsto R(\boldsymbol{c})$ \cite{QCRL_2024}.

Studies on QCRL are aimed at obtaining control pulses that achieve high robustness against noise while implementing different desired quantum operations.
One of the key applications of QCRL is an algorithm called RIPV (robustness-invariant pulse variation), whose key insight is to traverse the level set of QCRL to keep the robustness measure (denoted by $R$) invariant while updating the control parameters for a gate family associated with a specific variable (denoted by $\mathcal{J}$).
The basic idea of RIPV is to calculate the gradient of $\mathcal{J}$ and project it onto the subspace orthogonal to the gradient of $R$ using Gram-Schmidt process, and then update the control parameters along this projected direction to ensure that $R$ remains invariant while $\mathcal{J}$ changes.
For compactness, we denote the gradients of the target variable and the robustness measure by $\mathbf{g}_{\mathcal J}:=\partial \mathcal{J}/\partial \boldsymbol{c}$ and $\mathbf{g}_{R}:=\partial R/\partial \boldsymbol{c}$, respectively.
The detailed procedure is summarized in Algorithm~\ref{alg:RIPV}.

\begin{center}
\begin{minipage}{0.95\linewidth}
\refstepcounter{algorithmctr}
\hrule\vspace{0.4em}
\textbf{Algorithm \thealgorithmctr.} RIPV algorithm\label{alg:RIPV}
\vspace{0.4em}\hrule\vspace{0.6em}
\begin{algorithmic}[1]

\Require Robustness measure $R$, initial parameter $\boldsymbol{c}_0$ of control pulses with corresponding initial continuous variable $\mathcal J_0$ and increment of variable $\Delta \mathcal{J}$.

\State $\boldsymbol{c}=\boldsymbol{c}_0,\mathcal{J}=\mathcal{J}_0$
\While {$\mathcal{J} < \mathcal{J}_{\rm target}$}
    \State Compute projected direction
    \Statex \hspace{\algorithmicindent} $\mathbf{g}_{\mathcal J}^{\perp}=\mathbf{g}_{\mathcal J}-\langle \mathbf{g}_{R},\mathbf{g}_{\mathcal J}\rangle\frac{\mathbf{g}_{R}}{\norm{\mathbf{g}_{R}}^2}$

    \State Compute increment
    \Statex \hspace{\algorithmicindent} $\Delta \boldsymbol{c}=\Delta \mathcal{J}\cdot\mathbf{g}_{\mathcal J}^{\perp}/\langle \mathbf{g}_{\mathcal J},\mathbf{g}_{\mathcal J}^{\perp}\rangle$

    \State Update $\boldsymbol{c}=\boldsymbol{c}+\Delta \boldsymbol{c}$
    
    \State Calculate $\mathcal{J}$ using newly updated $\boldsymbol{c}$
\EndWhile
\end{algorithmic}
\vspace{0.4em}\hrule
\end{minipage}
\end{center}

In particular, for the single-qubit $R_x(\theta)$ gate family, the variable $\mathcal{J}$ reduces to the scalar angle $\theta$.

\subsection{Filter Function} \label{sec:filter_function}

The filter function formalism offers valuable insights into quantifying the impact of stochastic time-varying noise on quantum gates. Following the formalism introduced in reference \cite{Qctrl_FF_2021}, the robustness metric in Eq.~\eqref{eq:robust_eq} can be approximated by
\begin{equation} \label{robustness_metric}
    \mathcal{I}_{\rm robust}\approx \frac1{2\pi}\sum_{k} \int_{\omega_{\rm{min}}}^{\omega_{\rm{max}}} {\rm d}\omega F_k(\omega)S_k(\omega),
\end{equation}
given the frequency domain of interest $[\omega_{\rm min}, \omega_{\rm max}]$. 
$S_k(\omega)$ is the $k$-th noise power spectrum, and the corresponding filter function $F_k(\omega)$ is defined as
\begin{equation}
    F_k(\omega)=\frac{1}{{\rm Tr}(P)}{\rm Tr}\left(P\mathcal{G}_k(\omega)\mathcal{G}_k^\dagger(\omega)\right)
\end{equation}
with $\mathcal{G}_k(\omega)$ representing the Fourier transform of the modified noisy Hamiltonian in the interaction picture $\widetilde H_{n,k}'(t)$:
\begin{equation}
    \mathcal{G}_k(\omega):=\mathcal{F}\{\widetilde H_{n,k}'(t)\}(-\omega).
\end{equation}
Here the Fourier transform is defined as $\mathcal{F}\{f(t)\}(\omega)=\int_{-\infty}^{+\infty} f(t)e^{-i\omega t} {\rm d}t$, and $\widetilde H_{n,k}'(t)$ is obtained from the interaction-picture noisy Hamiltonian $\widetilde H_{n,k}(t)$ via a gauge transformation that introduces only a global phase factor
\begin{equation}
    \widetilde H_{n,k}'(t) = \widetilde H_{n,k}(t)-\frac{{\rm Tr}(P\widetilde H_{n,k}(t))}{{\rm Tr}(P)}\mathbb{I}.
\end{equation}

\section{Filter Function-enhanced QCRL Algorithm}

In this section, we present the filter function-enhanced QCRL (FF-QCRL) framework in detail.
We consider quantum systems subject to stochastic time-varying noise, with dynamics governed by the total Hamiltonian in Eq.~\eqref{hamiltonian_total}.
Our objective is to design control pulses that implement a family of parametric quantum gates $U_{\rm target}(\theta)$ while maximizing robustness against the stochastic noise characterized by specific power spectral densities (PSD) $S_k(\omega)$.

To achieve this, we need to simultaneously consider the noise-free fidelity and the robustness against noise.
The noise-free fidelity can be quantified by the infidelity between the control propagator $U_{\rm ctrl}(T)$ and the target gate $U_{\rm target}(\theta)$, which can be expressed as
\begin{equation} \label{infidelity_parametric}
    \mathcal{I}_{\rm fidelity}(\theta)=1-\left|{\rm Tr}\left(U_{\rm target}^\dagger(\theta) U_{\rm ctrl}(T)\right) / d \right|^2,
\end{equation}
where $d$ is the dimension of the system.
For some parametric gates, the fidelity (or infidelity) can be simply characterized by the difference of the target parameter $\theta$ and the actual parameter $\theta_{\rm actual}$ realized by the control pulse.
To quantify robustness against stochastic time-varying noise, we employ the filter function formalism introduced in Section \ref{sec:filter_function}.
The robustness metric can then be expressed as a function of the filter functions and the noise power spectral densities (PSD), which is demonstrated in Eq.~\eqref{robustness_metric}.

Our method contains two parts. The first part (see Section~\ref{sec:initialization}) optimizes the control pulse of a gate associated with a parameter $\theta_0$ by minimizing both the infidelity in Eq.~\eqref{infidelity_parametric} and the robustness metric in Eq.~\eqref{robustness_metric}, which together define a multi-objective optimization problem. The second part (see Section~\ref{sec:generation}) uses the QCRL framework to efficiently generate control pulses for different target parameters $\theta$ from the initial control pulse while preserving both high fidelity and robustness.

\subsection{Control Pulse Parameterization}

To facilitate the process of the following stages, we parameterize the control pulses $u_j(t)$ using a vector of parameters $\boldsymbol{c}=\left(c_1,c_2,\dots,c_P\right)^{\rm T}$.

This parameterization can take various forms, such as piecewise-constant functions, Fourier series expansions, or other suitable basis functions that capture the essential features of the control pulses while adhering to experimental constraints like bandwidth and maximum amplitude. For example, using a Fourier basis expansion, the control pulse parameterized by $\boldsymbol{c} = \left(a_0,a_1,\dots,a_N, b_1, \dots, b_N\right)^{\rm T}$ can be expressed as
\begin{equation}
    \begin{aligned}
        u(t) = W(t) \bigg(\sum_{l=0}^N & a_l \cos\left(\frac{2l\pi t}{T}\right) \\
        & + \sum_{l=1}^N b_l \sin\left(\frac{2l\pi t}{T}\right) \bigg),
    \end{aligned}
\end{equation}
where $W(t)$ is a window function that enforces boundary conditions on the control pulse, such as zero amplitude at the temporal endpoints.

\subsection{Initialization of Robust Control Pulses} \label{sec:initialization}

At this stage, our goal is to optimize the control pulse for a specific quantum gate parameterized by $\theta_0$. Specifically, we minimize both the infidelity in Eq.~\eqref{infidelity_parametric}, which enforces $U_{\rm ctrl}(T) \approx U_{\rm target}(\theta_0)$, and the robustness metric in Eq.~\eqref{robustness_metric}, which enforces $\tilde{U}_{\rm noise}(T) \approx \mathbb{I}$. We formulate this task as a multi-objective optimization problem, where the objective is to find the control pulse $u_j(t)$ that achieves the best trade-off between fidelity and robustness.

Our cost function is designed to balance these two objectives, along with any additional constraints that may be relevant to the control pulse design, such as power limitations or smoothness requirements.

The overall cost function can be formulated as
\begin{equation} \label{overall_cost_function}
    \begin{aligned}
        \mathcal{L}_{\rm cost} = &\lambda_1 \mathcal{L}_{\rm fidelity} + \lambda_2 \mathcal{L}_{\rm robust}\\
        & + \lambda_3 \mathcal{L}_{\rm amp} + \lambda_4 \mathcal{L}_{\rm smooth},
    \end{aligned}
\end{equation}
where
\begin{gather}
    \mathcal{L}_{\rm fidelity} = 1-\left|{\rm Tr}\left(U_{\rm target}^\dagger(\theta_0) U_{\rm ctrl}(T)\right) / d \right|^2, \label{F_fidelity_version_1} \\
    \mathcal{L}_{\rm robust} = \frac1{2\pi}\sum_{k=1}^{N_n} \int_{\omega \in \mathcal{A}_k} {\rm d}\omega F_k(\omega)S_k(\omega), \label{F_robustness_metric} \\
    \mathcal{L}_{\rm amp} = \sum_{j=1}^{N_u} \int_0^T |u_j(t)|^2 {\rm d}t, \label{F_amplitude_constraint} \\
    \mathcal{L}_{\rm smooth} =  \sum_{j=1}^{N_u} \int_0^T \left|\frac{{\rm d}u_j(t)}{{\rm d}t}\right|^2 {\rm d}t. \label{F_smoothness_constraint}
\end{gather}
Here, $\mathcal{A}_k \subseteq [\omega_{\min},\omega_{\max}]$ denotes the selected target band for the $k$-th noise channel, and $\lambda_{1,2,3,4}$ are weighting coefficients that balance the relative importance of each objective.
The terms $\mathcal{L}_{\rm fidelity}$ and $\mathcal{L}_{\rm robust}$ encode our dual optimization objectives, namely minimizing gate infidelity while maximizing noise resilience.
The penalty terms $\mathcal{L}_{\rm amp}$ and $\mathcal{L}_{\rm smooth}$ enforce constraints on the amplitude and smoothness of the control pulse.
For certain parametric gates where the parameter $\boldsymbol{\theta}$ is directly related to the control pulse—such as single-qubit rotation gates driven by one control field satisfying $\theta_{\rm actual} = \int_0^T u(t) {\rm d}t$—the fidelity term can be simplified to:
\begin{equation} \label{F_fidelity_version_2}
    \mathcal{L}_{\rm fidelity} = (\theta_{\rm actual} - \theta_0)^2.
\end{equation}

To validate the initialization stage, we will compare the optimized pulse with baseline pulses through: (1) control waveform comparison, (2) filter function analysis to assess noise suppression effectiveness, and (3) Monte Carlo simulations to evaluate fidelity under stochastic noise.

\subsection{Generation of Robust Control Pulses} \label{sec:generation}

From the past stage, we obtain an optimal control pulse $u_j^{\rm init}(t)$ parameterized by $\boldsymbol{c}_0$ for the gate associated with the parameter $\boldsymbol{\theta}_0$.

At this stage, we utilize the QCRL framework to efficiently find control pulses for different target values of $\boldsymbol{\theta}$ from the initial control pulse while preserving both high fidelity and robustness. This involves integrating the QCRL approach with the filter function formalism to adapt the control pulses.

We employ the RIPV algorithm (see Algorithm~\ref{alg:RIPV}) to traverse the level set of the robustness metric while updating the control parameters to achieve the desired target parameter $\boldsymbol{\theta}$, where the robustness metric $R = \mathcal{L}_{\rm robust}$ is defined in Eq.~\eqref{F_robustness_metric}.

To validate the generation stage, we will: (1) plot the evolution of pulse parameters $\boldsymbol{c}$ as a function of the target parameter $\boldsymbol{\theta}$, where smooth variations confirm continuous differentiability essential for interpolation, and (2) perform Monte Carlo simulations to verify that the generated pulses maintain robustness across the parameter range.

\section{Variational Quantum Circuit Surrogate}\label{sec:vqc_surrogate}

Although the FF-QCRL framework developed above can generate robust pulse families under stochastic time-varying noise, its practical implementation can still be computationally demanding in the generation stage.
In particular, the RIPV iteration repeatedly evaluates the robustness metric $\mathcal{L}_{\rm robust}$ and its gradient with respect to the pulse parameters.
When $\mathcal{L}_{\rm robust}$ is computed through filter-function integrals over multiple frequency bands, this repeated evaluation becomes a non-negligible computational bottleneck.
To explore an auxiliary quantum-assisted route for repeated robustness evaluation, we further introduce a variational quantum circuit (VQC) as a black-box surrogate model for robustness evaluation, following the general surrogate-modelling motivation in recent quantum machine learning benchmarks \cite{icaart24}.
This auxiliary module does not modify the FF-QCRL formulation itself, but replaces the robustness-related evaluation block in RIPV.

Let $\boldsymbol{c}=\left(c_1,c_2,\dots,c_P\right)^{\rm T}$ denote the parameter vector of the control pulse.
In the single-qubit, single-noise setting considered in this work, the pulse is parameterized by $P=7$ variables.
The purpose of the VQC is to approximate the mapping from the pulse parameters to the robustness metric $\mathcal{L}_{\rm robust}$ introduced in Eq.~\eqref{F_robustness_metric}, that is,
\begin{equation}
    \boldsymbol{c} \mapsto \hat{\mathcal{L}}_{\rm robust}(\boldsymbol{c};\varphi),
\end{equation}
where $\varphi$ denotes the trainable parameters of the circuit.
To make the surrogate specification explicit while keeping the module lightweight, we instantiate the predictor as
\begin{equation}
    \hat{\mathcal{L}}_{\rm robust}(\boldsymbol{c};\varphi)=g\!\left(\left\langle O\right\rangle_{U_{\rm VQC}\left(x(\boldsymbol{c}),\varphi\right)}\right),
\end{equation}
where $x(\boldsymbol{c})$ denotes angle encoding of the input parameters, $U_{\rm VQC}(\cdot,\varphi)$ is an $L$-layer variational circuit composed of parameterized single-qubit rotations and entangling gates, $O$ is the measured observable, and $g(\cdot)$ is a scalar classical post-processing map.

\begin{figure*}[t]
    \centering
    \includegraphics[width=0.95\textwidth]{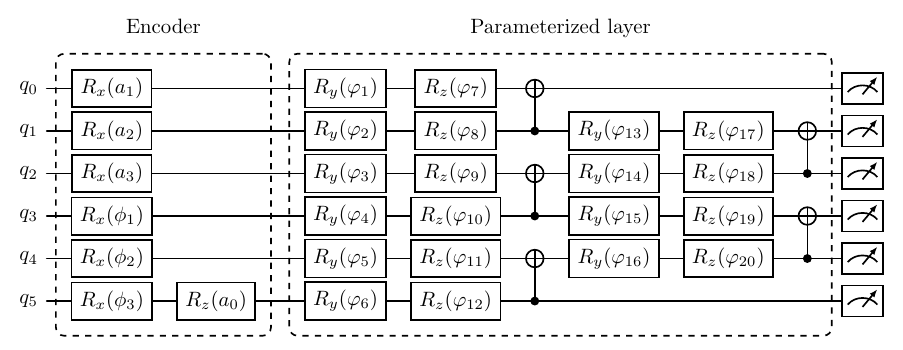}
\caption{The VQC surrogate architecture for RIPV. Pulse parameters are encoded into a variational quantum circuit and measured through an observable readout with classical post-processing. The resulting surrogate output is differentiated with respect to input parameters and used as an auxiliary update-direction signal in RIPV.}
\label{fig:vqc_architecture}
\end{figure*}

To keep the method portable across different control settings, we use a black-box PQC design rather than hand-crafting a circuit from the physical Hamiltonian.
The concrete implementation in this paper is shown in Figure~\ref{fig:vqc_architecture}: a 6-qubit, 2-block hierarchical-bridge CNOT ansatz with linear-$Z$ expectation readout and a shallow classical post-processing head, implemented in MindSpore Quantum \cite{xu2024mindspore}. Before encoding, input parameters are normalized to $[-\pi,\pi]$ (amplitude rescaled, phase already in range), and the VQC consists of two stacked encoder-plus-parameterized-layer blocks. This design is consistent with data re-uploading style PQC modeling and expressibility-oriented circuit design \cite{PerezSalinas2020datareuploading,Sim2019Expressibility,icaart24,PhysRevApplied.10.024052}.
Since the surrogate depends on pulse parameters rather than directly on the target rotation angle, one trained model can be reused across different target angles in the present setting.

In the current implementation, the surrogate is trained only with the exact values of $\mathcal{L}_{\rm robust}$ computed from the classical FF-QCRL model.
The gradient required by RIPV is then obtained by differentiating the trained surrogate output with respect to the input parameters, so that $\nabla_{\boldsymbol{c}}\hat{\mathcal{L}}_{\rm robust}(\boldsymbol{c};\varphi)$ serves as an approximation of $\nabla_{\boldsymbol{c}}\mathcal{L}_{\rm robust}(\boldsymbol{c})$.
More explicitly, the deployment gradient follows the input chain
\begin{equation}
    \nabla_{\boldsymbol{c}} \hat{\mathcal{L}}_{\rm robust}(\boldsymbol{c};\varphi)=\frac{\partial \hat{\mathcal{L}}_{\rm robust}}{\partial x}\frac{\partial x}{\partial \boldsymbol{c}},
\end{equation}
which is obtained by differentiating the trained surrogate with respect to the input parameters and then used in RIPV.

The training data for the surrogate are generated offline from the classical FF-QCRL model.
For each sampled pulse parameter vector $\boldsymbol{c}_i$, the exact robustness metric $\mathcal{L}_{\rm robust}(\boldsymbol{c}_i)$ is computed and used as the supervision label.
Accordingly, a representative value-based training objective is given by
\begin{equation}
    L(\varphi)=\frac{1}{N_s}\sum_{i=1}^{N_s}\left|\hat{\mathcal{L}}_{\rm robust}(\boldsymbol{c}_i;\varphi)-\mathcal{L}_{\rm robust}(\boldsymbol{c}_i)\right|^2.
\end{equation}
This choice keeps the training objective simple and avoids the additional cost of preparing explicit gradient labels, while still allowing the surrogate gradient to be extracted during deployment.

After training, the exact robustness-related evaluations in RIPV are replaced by the surrogate approximation
\begin{equation}
    R(\boldsymbol{c})=\mathcal{L}_{\rm robust}(\boldsymbol{c})\approx \hat{\mathcal{L}}_{\rm robust}(\boldsymbol{c};\varphi).
\end{equation}
Accordingly, the gradient $\nabla_{\boldsymbol{c}}\hat{\mathcal{L}}_{\rm robust}(\boldsymbol{c};\varphi)$ is used in the Gram--Schmidt projection step of Algorithm~\ref{alg:RIPV}.
By contrast, the quantity associated with the target gate rotation angle is still computed classically. Therefore, the VQC module is a local replacement of the robustness-evaluation block rather than a full rewriting of RIPV. The practical criterion is whether this replacement preserves downstream pulse-family robustness and provides useful update-direction information. In the present low-dimensional case, this module should be interpreted as a proof-of-concept exploration of quantum-assisted representation, not as a claim of demonstrated computational speedup.

\section{Numerical Results}

The numerical studies reported in this section were carried out using a hybrid software stack. In particular, the quantum-circuit components were implemented using MindSpore Quantum \cite{xu2024mindspore}, while the classical numerical routines for control optimization, filter-function evaluation, and Monte Carlo simulation were implemented using standard scientific-computing tools including JAX \cite{jax2018github}. The source code and data required to reproduce the reported numerical results are publicly available in a GitHub repository \cite{github_repo}.

\subsection{Robust single-qubit gate against single-source noise}

\begin{figure}[tbp]
    \centering
    \subfloat[]{
        \includegraphics[width=\linewidth]{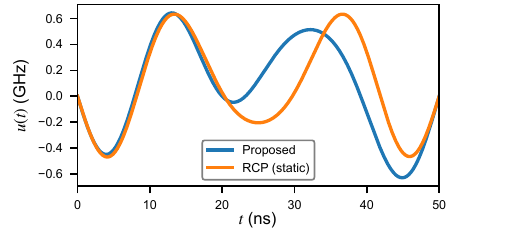}
        \label{fig:initial_pulse_comparison}
    }\\[0.5em]
    \subfloat[]{
        \includegraphics[width=\linewidth]{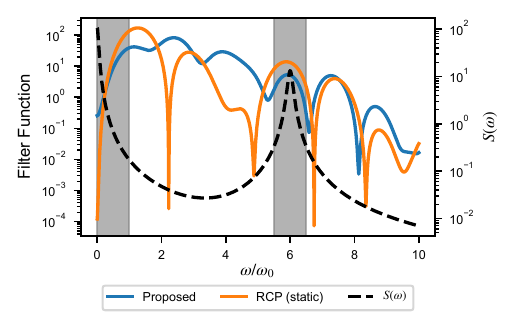}
        \label{fig:initial_filter_function_comparison}
    }
\caption{Initialization results for the gate $R_x(\pi)$ under stochastic detuning noise. (a) Comparison of the waveforms between the initial guess (RCP (static)) and the optimized pulse (Proposed). (b) Filter functions on logarithmic scale with the noise PSD $S(\omega)$ (dashed line) and the gray-shaded regions indicating the targeted noise bands.}
\label{fig:demonstration_of_initialization}
\end{figure}

In this section, we provide a proof-of-principle validation of the proposed method on the $R_x(\theta)$ gate under stochastic detuning noise. The total Hamiltonian of the system (see Eq.~\eqref{hamiltonian_total}) is given by

\begin{equation}
    H_{\rm tot}=\frac{\delta(t)}2\sigma_z+\frac{\Omega(t)}2\sigma_x,
\end{equation}

where $\delta(t)$ is the stochastic detuning noise, $\Omega(t)$ is the control pulse, and $\sigma_{x,z}$ are the Pauli matrices.
The target operator is $U_{\rm target}=R_x(\theta)=\exp(-i\frac\theta2\sigma_x)$ with the rotation angle $\theta\in[0,2\pi]$.
In the numerical example, we adopt the amplitude-phase form of the truncated Fourier expansion.
The control pulse $\Omega(t)$ is parameterized by $\boldsymbol{c} = \left(a_0, a_1, \ldots, a_N, \phi_1, \ldots, \phi_N\right)$ using a Fourier basis expansion as
\begin{equation}
    \Omega(t) = \sin\left(\frac{\pi t}T\right)\left(a_0+\sum_{l=1}^Na_l\cos\left(\frac{2l\pi t}T + \phi_l\right)\right),
\end{equation}
where $T$ is the pulse duration, and $N$ is the number of Fourier components.
We use $T = 50 \mathrm{ns}$ (corresponding to $\omega_0 = 2\pi/T \approx 0.126 \, \mathrm{GHz}$) and $N=3$ in the simulation.
Note that the window function $\sin\left(\frac{\pi t}T\right)$ can enforce $\Omega(t)\rightarrow 0$ when $t\rightarrow 0$ or $t\rightarrow T$.
The stochastic detuning noise $\delta(t)$ is modeled as multi-band colored noise, whose power spectral density (PSD) is shown in Figure~\ref{fig:initial_filter_function_comparison}.

First, we initialize the control pulse for the $R_x(\pi)$ gate using the method described in Section~\ref{sec:initialization}.
The cost function is defined by Eq.~\eqref{overall_cost_function} and Eqs.~\eqref{F_robustness_metric}-\eqref{F_fidelity_version_2} with weights $\lambda_1=1$, $\lambda_2=0.03$, $\lambda_3=10^{-4}$, and $\lambda_4=10^{-4}$.
Here, $S_k(\omega)$ in Eq.~\eqref{F_robustness_metric} is normalized with $\int_{-\infty}^{\infty} S_k(\omega) d\omega = 1$, and we target two noise bands: low frequencies $(0, \omega_0)$ and high frequencies $(5.5 \omega_0, 6.5 \omega_0)$.

We choose the control pulse with the parameters given by $R^{\pi}_{\mathrm{ex};\perp}$ from \cite{PhysRevApplied.23.054002}, which is a robust control pulse (RCP) optimized for robustness against quasi-static noises, as the initial guess for the optimization. Adopting the same control model and a closely related baseline enables a controlled comparison between robustness designed for quasi-static noise and robustness designed for stochastic time-varying noise. We then compare the optimized pulse with this baseline pulse and a simple sine pulse under stochastic detuning noise.

To quantitatively characterize the impact of noise on control performance and demonstrate the robustness improvement achieved by the proposed method, we employ Monte Carlo simulations to evaluate the average fidelity of different control pulses under stochastic noise.
In the following, $\delta_{\rm rms}$ denotes the root-mean-square (RMS) strength of stochastic noise, $\overline{\delta_{\rm rms}}$ denotes the sample mean of $\delta_{\rm rms}$ across multiple Monte Carlo samples.
The fidelity is defined as $\mathcal{F} = 1 - \mathcal{I}_{\rm fidelity}$ where $\mathcal{I}_{\rm fidelity}$ is defined in Eq.~\eqref{infidelity_parametric}, and $\langle \mathcal{F} \rangle$ denotes the average fidelity over multiple Monte Carlo samples.
By comparing the average fidelity of different control pulses at different $\overline{\delta_{\rm rms}}$ levels, we can clearly demonstrate the advantages of the proposed method in enhancing robustness.

Figure~\ref{fig:initial_pulse_comparison} compares the control waveforms of the initial guess (RCP (static), optimized for quasi-static noise) and the pulse optimized by the proposed method.
It can be seen that the optimization of the initialization stage achieves improved robustness (as validated below) without introducing significant changes in amplitude or smoothness, demonstrating the efficiency of the proposed approach.

Figure~\ref{fig:initial_filter_function_comparison} presents a comparison of the filter functions on a logarithmic scale, with the noise power spectral density $S(\omega)$ shown as a dashed line for reference.
The two gray-shaded regions indicate the targeted noise frequency bands: the low-frequency band $(0, \omega_0)$ and the high-frequency band $(5.5\omega_0, 6.5\omega_0)$.
In the low-frequency band, both pulses exhibit comparable filter function values, as expected since the RCP (static) baseline was originally designed to suppress quasi-static (i.e., near-zero frequency) noise.
However, substantial differences emerge in the higher frequency regions.
In the high-frequency band around $6\omega_0$, where the noise PSD exhibits a secondary peak, the optimized pulse maintains substantially lower filter function values.
This frequency-selective suppression directly translates to reduced noise sensitivity according to Eq.~\eqref{robustness_metric}, demonstrating the effectiveness of integrating the filter function formalism into the optimization framework.

\begin{figure}[tbp]
    \centering
    \subfloat[]{
        \includegraphics[width=0.95\linewidth]{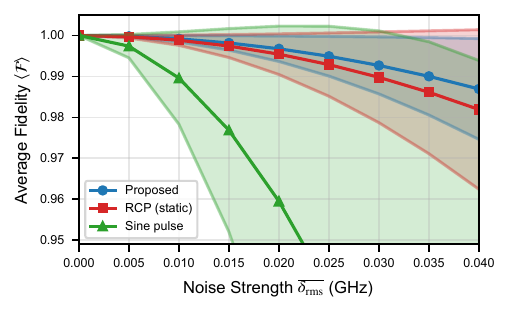}
        \label{fig:fidelity_comparison_initial}
    }\\[0.5em]
    \subfloat[]{
        \includegraphics[width=0.95\linewidth]{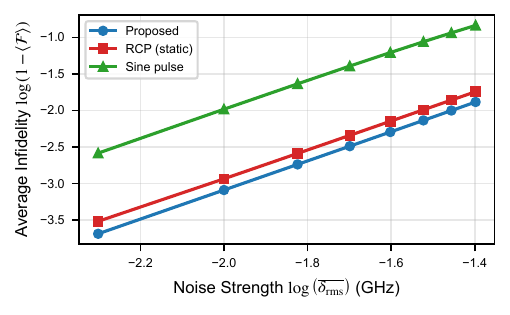}
        \label{fig:infidelity_comparison_initial}
    }
\caption{Monte Carlo validation of the initialization for the $R_x(\pi)$ gate under stochastic detuning noise. (a) Average fidelity versus RMS noise strength; shaded regions indicate $\pm 1$ standard deviation. (b) Average infidelity on logarithmic scale.}
\label{fig:demonstration2_of_initialization}
\end{figure}

Figure~\ref{fig:fidelity_comparison_initial} presents the fidelity comparison under stochastic detuning noise, where the horizontal axis represents the sample mean of the root-mean-square (RMS) noise strength $\overline{\delta_{\rm rms}}$.
The proposed method achieves consistently higher average fidelity compared to both baselines across the entire noise strength range.
In particular, while the sine pulse exhibits rapid fidelity degradation (dropping below 0.95 at $\overline{\delta_{\rm rms}} \approx 0.02$~GHz), both robust control pulses maintain fidelity above 0.98 even at $\overline{\delta_{\rm rms}} = 0.04$~GHz.
The shaded regions indicate the standard deviation of fidelity across 500 Monte Carlo realizations, demonstrating that the proposed method also exhibits reduced variance compared to the baselines, indicating more consistent performance under stochastic noise.

Figure~\ref{fig:infidelity_comparison_initial} shows the infidelity comparison on a logarithmic scale.
All three pulses exhibit approximately linear relationships between $\log(1-\langle \mathcal{F} \rangle)$ and $\log(\overline{\delta_{\rm rms}})$, with slopes close to 2, confirming that the average infidelity scales quadratically with noise strength, consistent with the observations of \cite{Zhang_2025}.
Crucially, the proposed method maintains a lower intercept than the RCP (static) throughout the tested range, with an improvement of approximately $30\%$ infidelity reduction.
The sine pulse, lacking robustness optimization, exhibits significantly higher infidelity (approximately one order of magnitude worse than the robust pulses).

Then, we generate control pulses for different rotation angles $\theta$ from the initial control pulse using the RIPV algorithm with the robustness metric $R = \mathcal{L}_{\rm robust}$.
The rotation angle $\theta$ varies from $\pi$ to $2\pi$ in increments of $\Delta \theta = 0.002~{\rm rad}$.

\begin{figure}[tbp]
    \centering
    \includegraphics[width=\linewidth]{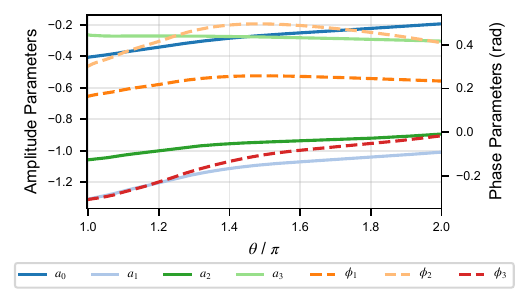}
\caption{Control pulse parameters for the $R_x(\theta)$ gate as functions of rotation angle $\theta \in [\pi, 2\pi]$.}
\label{fig:generation_pulse_parameters}
\end{figure}

Figure~\ref{fig:generation_pulse_parameters} shows the evolution of control pulse parameters as functions of $\theta$, where the smooth variations confirm the continuous differentiability essential for interpolation.
This enables us to obtain a continuous family of control pulses for the $R_x(\theta)$ gate with $\theta \in [\pi, 2\pi]$.

To validate robustness preservation across the gate family, we select 11 uniformly spaced rotation angles from $\pi$ to $2\pi$ and perform Monte Carlo simulations for each.

\begin{figure}[tbp]
    \centering
    \subfloat[]{
        \includegraphics[width=0.95\linewidth]{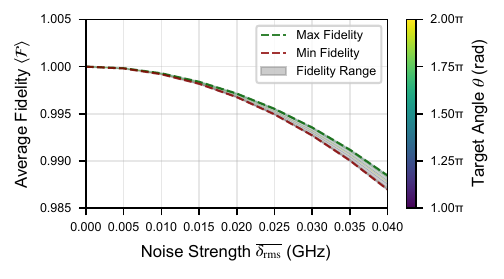}
        \label{fig:fidelity_comparison_generated}
    }\\[0.5em]
    \subfloat[]{
        \includegraphics[width=0.95\linewidth]{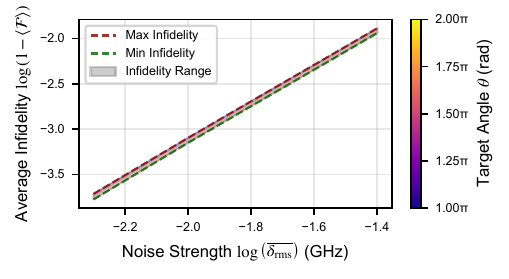}
        \label{fig:infidelity_comparison_generated}
    }
\caption{Monte Carlo validation of the generation for the gates $R_x(\theta)$ under stochastic detuning noise. (a) Average fidelity comparison under stochastic detuning noise. (b) Average infidelity comparison on a logarithmic scale under stochastic detuning noise.}
\label{fig:demonstration_of_generation}
\end{figure}

As shown in Figure~\ref{fig:fidelity_comparison_generated}, all the pulses generated maintain a high value of average fidelity exceeding $98.5 \%$ throughout the parameter range $\theta \in [\pi, 2\pi]$, with the gray-shaded region between the maximum and minimum fidelity curves demonstrating a negligible spread across different target angles.
The corresponding logarithmic-scale infidelity plot in Figure~\ref{fig:infidelity_comparison_generated} reveals that all pulses exhibit linear scaling with slope approximately equal to 2, consistent with the expected quadratic dependence on noise strength.
Notably, the infidelity remains tightly bounded between approximately $10^{-3.7}$ and $10^{-1.9}$ across noise strengths from $\overline{\delta_{\rm rms}} = 0.005$ to $0.04$~GHz, with the spread between maximum and minimum values remaining nearly constant regardless of $\theta$.
This uniformity across the gate parameter space demonstrates that the generation stage successfully preserves the superior robustness characteristics of the initial pulse throughout the entire gate family.

Overall, these results provide numerical evidence that the proposed FF-QCRL framework can generate robust parametric quantum gates under stochastic time-varying noise in the present single-qubit setting.

\subsection{VQC-assisted pulse-family generation}\label{sec:vqc_numerics}

We further evaluate the VQC surrogate introduced in Section~\ref{sec:vqc_surrogate} in the same single-qubit, single-noise setting considered above.
The control pulse is still parameterized by seven Fourier parameters, and the surrogate is trained offline using the exact values of $\mathcal{L}_{\rm robust}$ generated from the classical FF-QCRL model.
During deployment, robustness-related evaluations in RIPV are fully replaced by the surrogate, whereas the target-angle-related quantity is still computed classically.
For reproducibility, we report a minimal training configuration for the surrogate, including dataset size and split, optimizer and learning-rate schedule, training epochs, and early-stopping rule.

In the current setting, the dataset sampling cap is 512, with train/validation/test split 409/51/52, the optimizer is Adam with learning rate 0.01 and batch size 8, and stabilization includes gradient clipping (1.0) and early stopping (patience 4). The final number of training epochs is 17.

The validation of the VQC module in this work is task-oriented.
Rather than prioritizing pointwise function-fitting accuracy, we focus on whether the surrogate-assisted RIPV can improve downstream pulse-family performance and whether the surrogate provides non-random update-direction information in the RIPV step.
Accordingly, we report two groups of results: downstream gate performance under stochastic noise and ablation tests on update-direction quality \cite{icaart24}.

\begin{figure}[tbp]
    \centering
    \includegraphics[width=\linewidth]{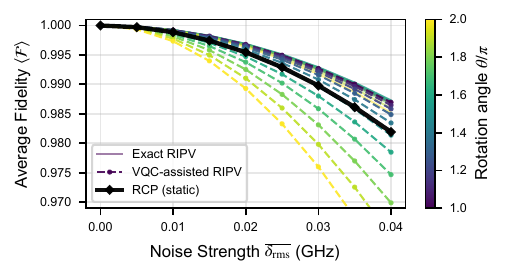}
\caption{Average fidelity versus noise strength in a single panel for target angles $\theta \in \{\pi,1.1\pi,\ldots,2\pi\}$. Colors encode $\theta$ through a shared color bar. Exact RIPV and VQC-assisted RIPV are distinguished by line style, and the baseline initial guess (RCP (static)) is shown for reference.}
\label{fig:vqc_generation_performance}
\end{figure}

We first examine whether replacing the robustness-related evaluations in RIPV by the VQC surrogate can still generate a high-quality pulse family.
Figure~\ref{fig:vqc_generation_performance} reports the average gate fidelity as a function of noise strength, where both exact RIPV and VQC-assisted RIPV are shown for target angles from $\pi$ to $2\pi$ with step $0.1\pi$.
This visualization directly compares the two RIPV variants under the same noise levels across the full angle range and is used to identify angle-dependent performance regimes of the current surrogate.
From Figure~\ref{fig:vqc_generation_performance}, the VQC-assisted RIPV remains close to exact RIPV for $\theta\in[\pi,1.5\pi]$, indicating partial transferability of the learned robustness signal in this interval. Around $\theta\approx1.5\pi$, the VQC-assisted curves can drop below the baseline initial guess (RCP (static)) under comparable noise levels. For $\theta\in(1.5\pi,2\pi]$, the gap to exact RIPV widens further with increasing noise strength, revealing a clear angle-dependent degradation regime for the current surrogate.

\begin{figure}[tbp]
    \centering
    \includegraphics[width=\linewidth]{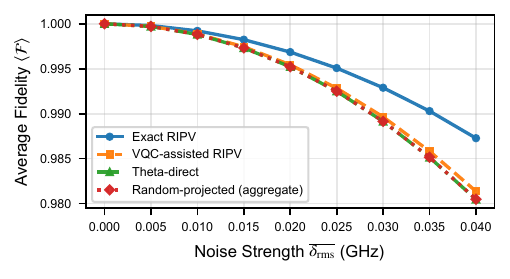}
\caption{Ablation of update-direction strategies for $R_x(1.5\pi)$ under a shared Monte Carlo noise batch. Average fidelity is plotted against noise strength $\overline{\delta_{\rm rms}}\in[0,0.04]$ for Exact RIPV, VQC-assisted RIPV, theta-direct update (without robustness orthogonalization), and random-projected update (aggregate over random substitute directions). The vertical axis is truncated to highlight small but systematic differences, which become more visible in the higher-noise regime.}
\label{fig:vqc_direction_ablation}
\end{figure}

To assess whether the surrogate provides useful directional information relative to heuristic alternatives, we conduct an ablation study for the target gate $R_x(1.5\pi)$ under a shared Monte Carlo noise batch.
Figure~\ref{fig:vqc_direction_ablation} plots the average fidelity as a function of noise strength $\overline{\delta_{\rm rms}}\in[0,0.04]$ for four update strategies: Exact RIPV, VQC-assisted RIPV, theta-direct update (i.e., direct angle-targeting update without robustness orthogonalization), and random-projected update (aggregated over random substitute directions).
All four strategies remain close in the low-noise regime, whereas their separation becomes progressively more visible as the noise strength increases. Exact RIPV remains consistently best, while VQC-assisted RIPV retains a small but systematic advantage over both heuristic baselines, indicating that the surrogate direction carries limited but non-zero task-relevant information.

To emphasize the practical high-noise regime, we further summarize the corresponding values at the maximum tested noise strength ($\overline{\delta_{\rm rms}}=0.04$~GHz) in Table~\ref{tab:vqc_ablation_summary}.
At this operating point, the ranking remains Exact RIPV $>$ VQC-assisted RIPV $>$ theta-direct $\approx$ random-projected, confirming that the surrogate-derived direction is beneficial but still falls short of Exact RIPV.

\begin{table}[htbp]
\centering
\caption{Ablation summary at the maximum tested noise strength, $\overline{\delta_{\rm rms}}=0.04$~GHz, extracted from Figure~\ref{fig:vqc_direction_ablation}.}
\label{tab:vqc_ablation_summary}
\begin{tabular}{l c}
\hline
Update strategy & Average fidelity \\
\hline
Exact RIPV & 0.98728 \\
VQC-assisted RIPV & 0.98137 \\
Theta-direct (no orthogonalization) & 0.98049 \\
Random-projected (aggregate) & 0.98047 \\
\hline
\end{tabular}
\end{table}

\section{Conclusion}

In this paper, we propose FF-QCRL (filter function-enhanced quantum control robustness landscape), which extends the quantum control robustness landscape (QCRL) framework to stochastic time-varying noise through integration of the filter function formalism. The resulting robustness metric quantifies gate sensitivity over relevant frequency bands and enables robust pulse design against spectrally complex noise.

The FF-QCRL approach consists of two stages: (1) an initialization stage that optimizes a control pulse for a specific gate parameter by minimizing both infidelity and the filter-function-based robustness metric, and (2) a generation stage that employs the RIPV algorithm to traverse the robustness landscape and produce a continuous family of control pulses for parametric quantum gates while preserving the optimized robustness. In the representative single-qubit setting studied here, numerical results for the gate $R_x(\theta)$ under multi-band stochastic detuning noise show an approximately $30\%$ infidelity reduction relative to pulses optimized only for quasi-static noise, confirming effective suppression of high-frequency components. The generated pulse family maintains high fidelity (exceeding $98.5\%$) with uniform robustness across $\theta \in [\pi, 2\pi]$.

To complement the classical FF-QCRL pipeline, we further introduce a VQC surrogate, implemented with MindSpore Quantum, as an auxiliary quantum-assisted route for approximating the robustness metric during RIPV-based pulse-family generation. This module preserves the FF-QCRL formulation while providing a reusable surrogate mechanism for repeated robustness-related evaluations.

Beyond the surrogate interpretation adopted in this work, the VQC component also hints at a possible Quantum-for-Quantum route for robust control design. More specifically, it may be possible to use a trainable quantum circuit built from individually non-robust pulse primitives to represent or guide the optimization of a family of robust pulses, in a way that is conceptually reminiscent of composite-pulse robustness emerging from structured composition. Investigating whether such a quantum-native ansatz can provide a genuine representational or computational benefit remains an open question for future work.

Future work includes extending the present validation to single-qubit settings with multiple noise channels and to two-qubit gates, completing and benchmarking the VQC-assisted implementation, investigating hybrid exact-correction strategies for long RIPV trajectories, and validating the overall approach experimentally on superconducting or trapped-ion processors.

\begin{acknowledgments}

The authors acknowledge the support of the National Natural Science Foundation of China (Grant Nos. 62273016 and 62173296), the support of the Guangdong Basic and Applied Basic Research Foundation (Grant No. 2025A1515010186), the support of the Guangdong Provincial Quantum Science Strategic Initiative (Grant No. GDZX2405005), and the partial support from the Science Center Program of the National Natural Science Foundation of China (Grant No. 62188101).
This work was sponsored by CPS-Yangtze Delta Region Industrial Innovation Center of Quantum and Information Technology-MindSpore Quantum Open Fund.

\end{acknowledgments}

\bibliography{ref}

\end{document}